\documentclass[reprint,superscriptaddress, nofootinbib,longbibliography]{revtex4-2}
\pdfoutput=1
\usepackage{graphicx}
\usepackage{amssymb}
\usepackage[linktoc=none]{hyperref}
\usepackage[dvipsnames]{xcolor}
\usepackage{dcolumn}
\usepackage[english]{babel}
\usepackage[utf8x]{inputenc}
\usepackage[T1]{fontenc}
\usepackage{eqnarray,amsmath,physics,bbold}
\usepackage{ dsfont }
\usepackage{appendix}
\usepackage{xcolor}
\usepackage{breqn}

\usepackage{hyperref}
    \hypersetup{
    colorlinks   = true,    % Colours links instead of ugly boxes
    urlcolor     = blue,    % Colour for external hyperlinks
    linkcolor    = blue,    % Colour of internal links
    citecolor    = red      % Colour of citations
    }

\DeclareMathAlphabet{\mathpzc}{OT1}{pzc}{m}{it}

\newcommand{\vek}{\Vec{k}}

\newcommand{\cdag}[1]{c_{#1}^\dagger}

\newcommand{\INFN}{INFN - Sezione di Napoli, Complesso Univ. Monte S. Angelo, I-80126 Napoli, Italy}
\newcommand{\UNINA}{Physics Department "Ettore Pancini", Universit\'a degli studi di Napoli "Federico II", Complesso Univ. Monte S. Angelo, Via Cintia, I-80126 Napoli, Italy}
\newcommand{\UNISA}{Physics Department "E.R. Caianiello", Universit\'a degli studi di Salerno, Via Giovanni Paolo II, 132, I-84084 Fisciano (Sa), Italy}
\newcommand{\CNR}{CNR-SPIN Napoli Unit, Complesso Univ. Monte S. Angelo, Via Cintia, I-80126 Napoli, Italy}

\begin{document}
\title{Tunable spin and orbital Edelstein effect at (111) LaAlO$_3$/SrTiO$_3$ interface}
%\author{Mattia Trama}

\author{M. Trama}
\email{mtrama@unisa.it}
\affiliation{\UNISA}
\affiliation{\INFN}

\author{V. Cataudella}
\affiliation{\UNINA}
\affiliation{\CNR}

\author{C. A. Perroni}
\affiliation{\UNINA}
\affiliation{\CNR}

\author{F. Romeo}
%\email{}
\affiliation{\UNISA}

\author{R. Citro}
\email{rocitro@unisa.it}
\affiliation{\UNISA}
\affiliation{\INFN}

\begin{abstract}
    Converting charge current into spin current is one of the main mechanisms exploited in spintronics. One prominent example is the Edelstein effect, namely the generation of a magnetization in response to an external electric field, which can be realized in systems with lack of inversion symmetry.
    %In a system with lack of inversion symmetries, a static electric field can generate a magnetization by Edelestein effect. 
    If a system has electrons with an orbital angular momentum character, an orbital magnetization can be generated by the applied electric field giving rise to the  so-called orbital Edelstein effect. Oxide heterostructures are the ideal platform for these effects due to the strong spin-orbit coupling and the lack of inversion symmetries. Beyond a gate-tunable spin Edelstein effect, we predict an {\it orbital} Edelstein effect an order of magnitude larger then the spin one at the (111) LaAlO$_3$/SrTiO$_3$ interface. We model the material as a bilayer of $t_{2g}$ orbitals using a tight-binding approach, while transport properties are obtained in the Boltzmann approach.
    We give an effective model at low filling which explains the non-trivial behaviour of the Edelstein response, showing that the hybridization between the electronic bands crucially impacts the Edelstein susceptibility. 
\end{abstract}

\maketitle
\noindent\textit{Introduction.}
The spintronics is an emergent field that exploits the intrinsic spin of the electrons, in addition to its charge.
The goal is to produce devices which combine information storage, sensing, and processing in a single platform. In view of their characteristics, these devices could in principle overcome the performances of standard electronic devices in terms of data processing speed and consumption~\cite{dieny2020opportunities}. A possibility for spin manipulation is the injection of spin current from ferromagnets to semiconductors, which is however inefficient~\cite{wolf2001spintronics}. 
\\The best option is provided by the spin-to-charge interconversion, which allows to generate spin current directly inside the materials. In non-magnetic systems this can be realized, either by the spin Hall effect or the Edelstein effect (EE). The former is the creation of a transverse spin current in response to a charge current~\cite{dyakonov1971current}, while the latter is the spin accumulation in response to an applied electric field~\cite{edelstein1990spin, aronov1989nuclear}.
%This spin accumulation can be explained as follows: assume we have two quadratic bands of a Kramers doublet which are non-degenerate due to the Rashba splitting. For each band the spin circles around the contour of the Fermi surface in different direction. In this situation the total spin is zero for each band. When an electric field $\vec{E}$ is added, the Fermi surfaces are shifted by a different amount proportional to $\vec{E}$. In this new situation the spin is non-zero for each band and the sum of the two bands is not balanced leading to the magnetization.
\begin{figure}
    \centering
    \includegraphics[width=0.48\textwidth]{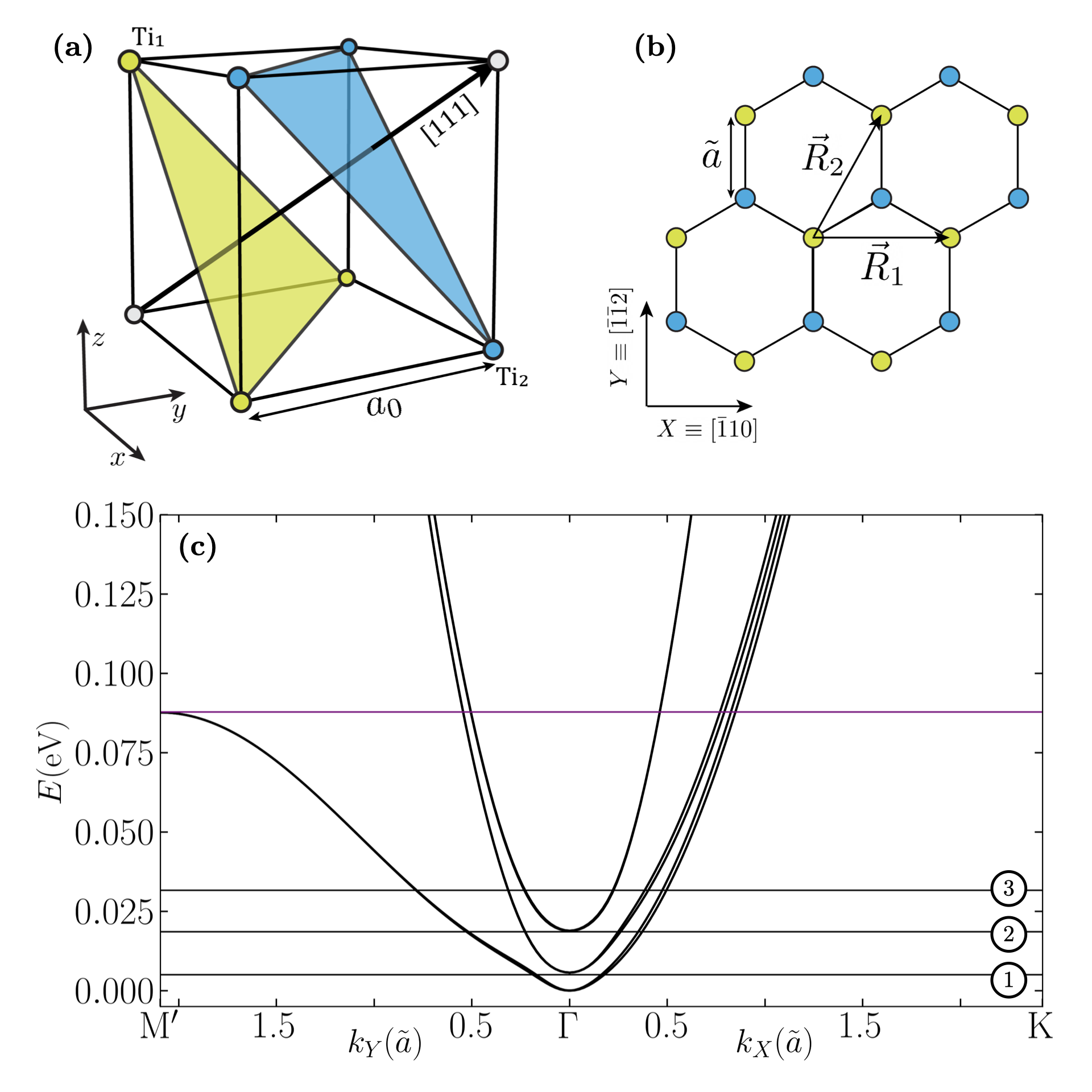}
    \caption{(a) Ti atoms in STO lattice, whose lattice constant is $a_0=0.3905$ nm. The orange and green dots represent atoms belonging to two non-equivalent planes.
    (b) Projection of the two non-equivalent planes of Ti over the (111) plane with our choice of primitive vectors $\Vec{R}_1$ and $\Vec{R}_2$ and $\Tilde{a}=\sqrt{2/3}a_0$. (c) Band structure along two different directions in the Brillouin zone. The purple benchmark line corresponds to a Lifshitz transition (see Supplementary material~\cite{supplementary}).}
    \label{band_structure}
\end{figure}
This spin accumulation can be explained as follows: an electric field $\vec{E}$ shifts the Fermi surfaces of the non-degenerate Kramers doublets, leading to an imbalance of spin and consequently a magnetization occurs.
\\A crucial role in obtaining the EE is the lack of inversion symmetry, which causes a Rashba spin-orbit coupling (SOC) locking the spin with the quasi-momentum of the electrons in a crystal. Oxide heterostructures are therefore the perfect environments for such a coupling.
\begin{figure*}
    \centering
    \includegraphics[width=0.85\textwidth]{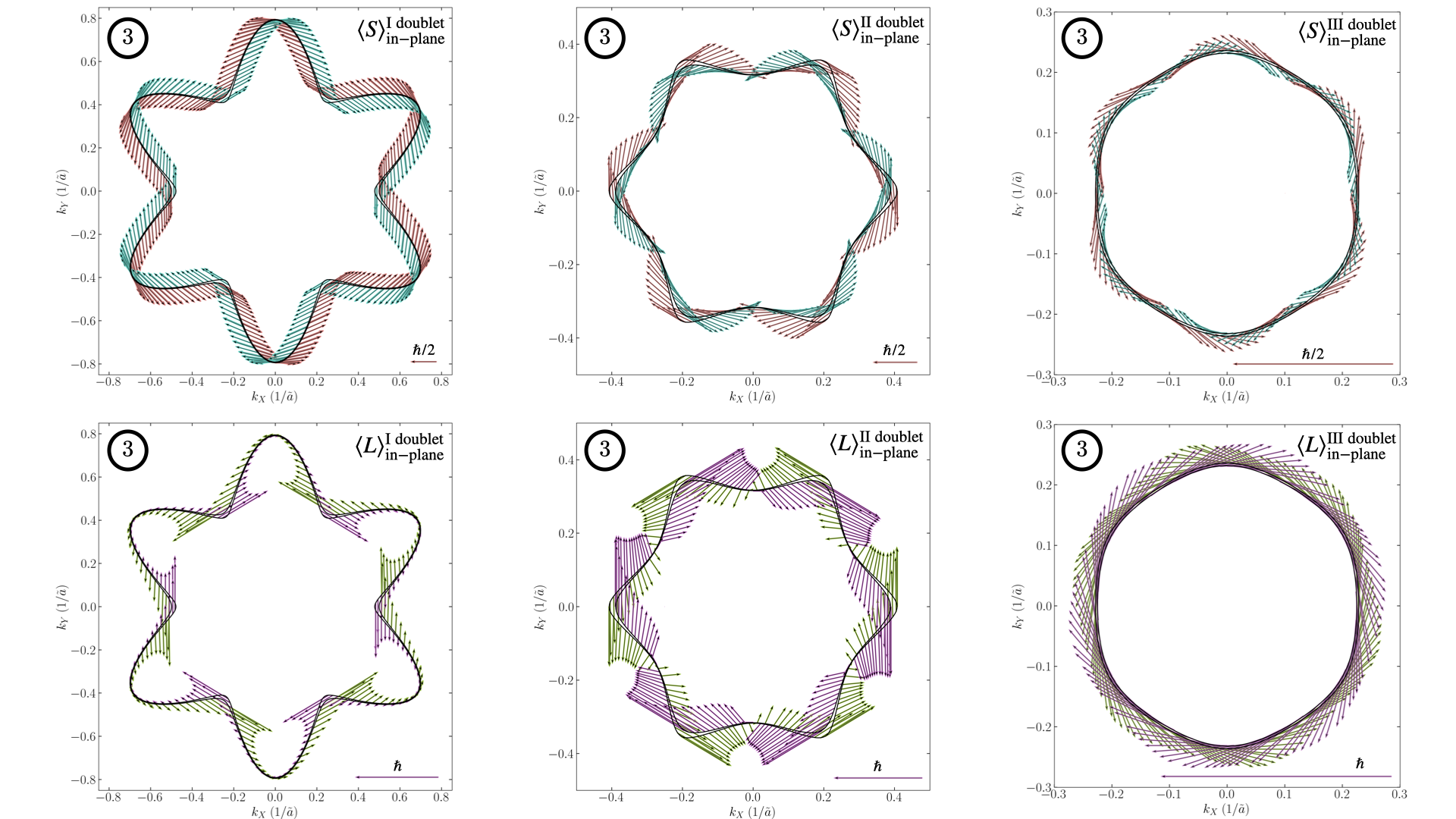}
    \caption{In-plane spin (upper panel) and orbital angular momentum (lower panel) textures for the three doublets with the chemical potential fixed to the value corresponding to the benchmark line 3 in Fig.~\ref{band_structure}. The red and green arrows represent the mean value of in-plane component of the operator for the external band, while the blue and pink refer to the internal component. The mean value of the generic operator $O$ is evaluated as $\langle O\rangle=\sqrt{\langle O_{\overline{1}10}\rangle^2+\langle O_{\overline{11}2}\rangle^2}$.}
    \label{Figura_spin}
\end{figure*}
In fact, the interface between the two insulating materials generates a quantum well for the electrons, forming a quasi-two dimensional electronic gas (2DEG), which naturally lays in a system with lack of inversion symmetry~\cite{lin2019interface}. Moreover, in these oxides the atomic SOC is typically stronger than in semiconductor interfaces due to the $d$ orbitals of the atom involved in the crystal structure~\cite{hwang2012emergent}.
Therefore, (001) SrTiO$_3$ (STO)-based heterostructures exhibit many non-trivial phenomena based on spin-orbital motion, such as tunable SOC~\cite{caviglia2010tunable}, generation and control of spin and orbital textures~\cite{gariglio2018spin}, coexistence of superconductivity and 2D magnetism~\cite{hwang2012emergent,pai2018physics} and topological properties both in normal and superconducting state~\cite{vivek_normal,scheurer2015topological,mohanta2014topological,loder2015route,fukaya2018interorbital,fidkowski2011majorana,fidkowski2013magnetic,mazziotti2018majorana,perroni2019evolution,perroni_ultimo}. Even if the inverse Edelstein effect, namely the generation of a charge current in response to a spin current, has been studied more extensively~\cite{trier2019electric}, only recently the EE has been taken into account in this system~\cite{Bibes_edelstein,chirolli2022colossal}. The results are promising, not only due to the presence of the canonical EE, but also for the presence of the so-called orbital Edelstein effect (OEE)~\cite{levitov1985magnetoelectric}, making this interface appealing for the field of orbitronics~\cite{revieworbitronic2021}. Since the electrons of 2DEG have a $d$ orbital character, an orbital magnetization occurs in response to an electric field.
\\The promising results obtained so far with (001) interfaces further motivates the interest into interfaces along other crystallographic directions. The (111) direction has been recently proven to be particularly promising, due to the hexagonal lattice of these structures, which is responsible for many non-trivial phenomena~\cite{chakhalian2020strongly,xiao2011interface,bruno2019band}. 
The (111) LaAlO$_3$/SrTiO$_3$ (LAO/STO) interface has been intensively studied~\cite{boudjada2017magnetic,rout2017six,monteiro2017two,davis2017magnetoresistance,doennig2013massive,khanna2019symmetry}. However, there are no predictions or experimental evidence on EE or OEE in this system, even though both the material and the direction are particularly interesting. In this system the strong orbital intermixing and the peculiar spin and orbital textures~\cite{he2018observation,trama2021straininduced,trama2022gate} suggest the possibility to establish an orbital magnetization, and could be of practical interest for the realization of spintronics and orbitronics devices. \\Therefore, in this work we theoretically predict the existence of the EE and OEE in the (111) LAO/STO interface, characterizing its properties. We model the material via a bilayer of Ti atoms using 3 orbital degrees of freedom treated by the tight binding (TB) approach, while the transport properties are modeled within the relaxation time approximation of the Boltzmann approach. We predict two different behaviours of the electrical response: a gate-tunable spin EE and an OEE an order of magnitude higher than the spin one, which cannot be explained in a common simplified Rashba model. We show that they emerge from the combined effect of the non-trivial Rashba SOC and the multi-orbital character of the electronic band structure.
%In this work we predict two different phenomena emerging from the combined effect of the non-trivial Rashba SOC and the multi-orbital character of the electronic band structure: a gate tunable \textit{spin} EE and a huge \textit{orbital} EE, which cannot be explained in a common simplified Rashba model.

\noindent\textit{Methods.}
The electronic band structure of LAO/STO interface can be obtained in terms of the $t_{2g}$ orbitals of the Ti atoms in STO \cite{Keppler1998}.
In order to take into account the electronic confinement, we use an accurate TB model, described in Ref.~\cite{trama2022gate}, of two layers of Ti atoms projected in the (111) direction, resulting in a honeycomb lattice as shown in Fig.~\ref{band_structure}. 
The Hamiltonian we take is
\begin{equation}
    H=H_{\text{TB}}(t_D,t_I)+H_{\text{SOC}}(\lambda)+H_{\text{TRI}}(\Delta)+H_{v}(v),
    \label{hamiltonian_comp}
\end{equation}
where $H_{\text{TB}}$ contains the direct and indirect first neighbour hopping terms, whose amplitude $t_D$ and $t_I$ are fixed in Ref.~\cite{trama2021straininduced} by fitting the angular resolved photoemission spectroscopy experimental data. $H_{\text{SOC}}$ is the atomic spin-orbit coupling of amplitude $\lambda=0.01$~eV~\cite{monteiro2019band} and $H_{\text{TRI}}$ is the trigonal cristal field~\cite{trama2021straininduced} of amplitude $\Delta=-0.005$~eV~\cite{de2018symmetry}. Finally, $H_{v}$ parametrizes the effect of the confinement which breaks the inversion symmetry and thus generates the so-called orbital Rashba~\cite{revieworbitronic2021}, whose amplitude depends on the electric potential $v$.
This term is responsible for the EE. In the region of low filling, a quadratic expansion in the quasimomentum $\vek$ of the Hamiltonian leads to the effective Hamiltonian~\cite{notaklim}
\begin{equation}
     H_{\rm{eff}}=\sum_{i=x,y,z}\mathcal{E}_i(\vek) (\mathbb{1}-L_i^2)-\frac{\lambda}{2} \hat{L}\cdot\hat{S}-\frac{3\Delta}{2} L_{111}^2+\mathcal{F}(\vek\times\hat{L})\cdot\hat{n}_{111}+\varepsilon_0,
     \label{lsmodel}
\end{equation}
where $\vek$ is expressed in units of the in-plane lattice constant $\Tilde{a}=\sqrt{2/3}a_0=\sqrt{2/3}\cdot0.3905$ nm, $\mathcal{E}_i$ is the renormalized dispersion expanded to second order at $\vek$:
\begin{align}
    &\mathcal{E}_x= 0.13 k_X^2 -0.29k_Xk_Y+0.29 k_Y^2,
    \label{secondorderTB1}
    \\&\mathcal{E}_y= 0.13 k_X^2+0.29 k_Xk_Y+0.29 k_Y^2,
    \label{secondorderTB2}
    \\&\mathcal{E}_z=0.37 k_X^2+0.044 k_Y^2,
    \label{secondorderTB3}
\end{align}
$L_i$ and $S_i$ are the $i-$th components of the orbital and spin angular momentum operator for $L=1$ and $S=1/2$, $L_{111}$ is the projection of the angular momentum along the $(111)$ direction, $\hat{n}_{111}$ is a unitary vector along the $(111)$ direction, the term $\vek\times\hat{L}$ is the orbital Rashba whose strength is included in the coefficient $\mathcal{F}=0.0035$~eV (depending on $v$ which is fixed to $0.2$~eV), and $\varepsilon_0$ is an energy constant.
The expressions and the numerical values of the coefficients in Eq.~(\ref{lsmodel}) can be found in the Supplementary material~\cite{supplementary}.
%The complete Hamiltonian~(\ref{hamiltonian_comp}), the derivation of Eq.~(\ref{lsmodel}), the expressions and the numerical value of the coefficients can be found in the Supplementary material of Ref.~\cite{trama2022gate}.
The combination of the atomic SOC and the orbital Rashba is translated into a generalized total angular momentum Rashba effect of the form $\hat{J}\times\vek$, where $\vec{J}$ is the total angular momentum. 
The electronic band structure in the low energy region is shown in Fig.~\ref{band_structure}. 
\begin{figure}
    \centering
    \includegraphics[width=0.45\textwidth]{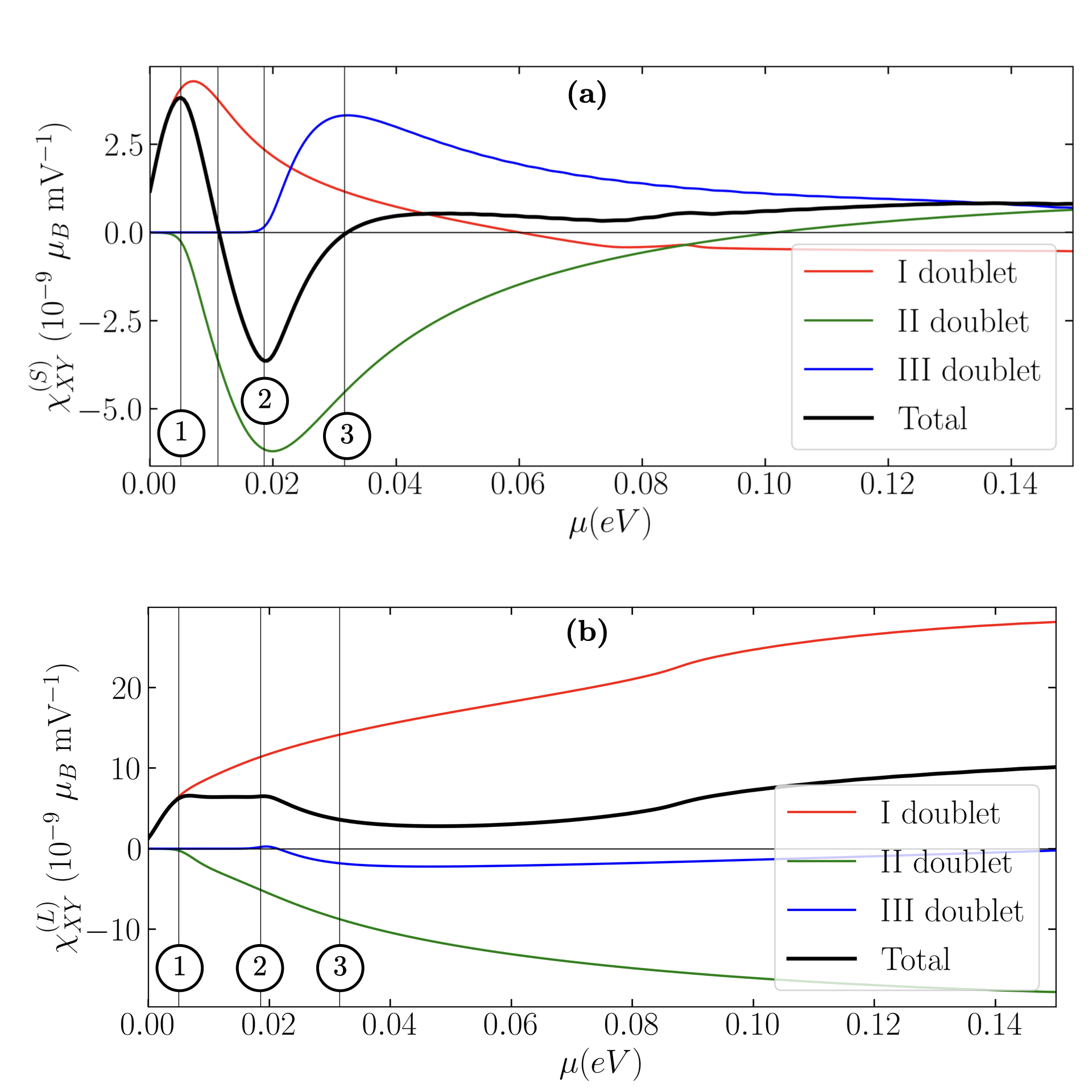}
    \caption{Spin (a) and orbital (b) Edelstein coefficient  as a function of the chemical potential. The different colors correspond to the contribution of a specific Kramers doublet.}
    \label{fig:Edelstein_chi}
\end{figure}
In absence of SOC and trigonal crystal field, all the bands would be degenerate in $\vek=0$.
The splitting between the doublets, due to these couplings, is smaller than in the most studied $(001)$ LAO/STO interface, which is crucial for the results we find. In fact the vicinity of the bands leads to a strong hybridization which amplifies the spin and oribital EE.
Near $\Gamma$ a linear Rashba splitting appears for the lowest Kramers doublet, while for the second doublet a cubic splitting in $\vek$ is found, differing from a simple description of a spin Rashba model~\cite{trama2022gate}. 
%This is due to $J_{111}$ character of the bands: the first and the third doublet are characterized by $J_{111}=1/2$ while the second doublet has $J_{111}=3/2$. Therefore, the Rashba interaction cannot split the bands within the second doublet since there are no $J_{111}=1/2$ states available, and a splitting occurs only at the third order of the perturbation theory. 
Far away from $\Gamma$, the $d_{xy},d_{yz},d_{zx}$ character of the bands is restored. The region in which the crossover between these two behaviors occurs is the most sensible to the hybridization of the bands.
By fixing the chemical potential to a benchmark value, we observe a non-trivial spin and orbital angular momentum texture on the Fermi surface in Fig.\ref{Figura_spin}.
First, both the spin and the orbital angular momentum are wrapping around the Fermi contour. 
%In Ref~\cite{trama2022gate} the out-of-plane component is shown.
The orbital pattern shows that the in-plane component is higher when the Fermi surfaces of two doublets are close to one another, pointing in the same direction, which is a sign of hybridization. The textures for the other benchmark lines are found in the supplementary material~\cite{supplementary}.
\begin{figure*}
    \centering
    \includegraphics[width=0.85\textwidth]{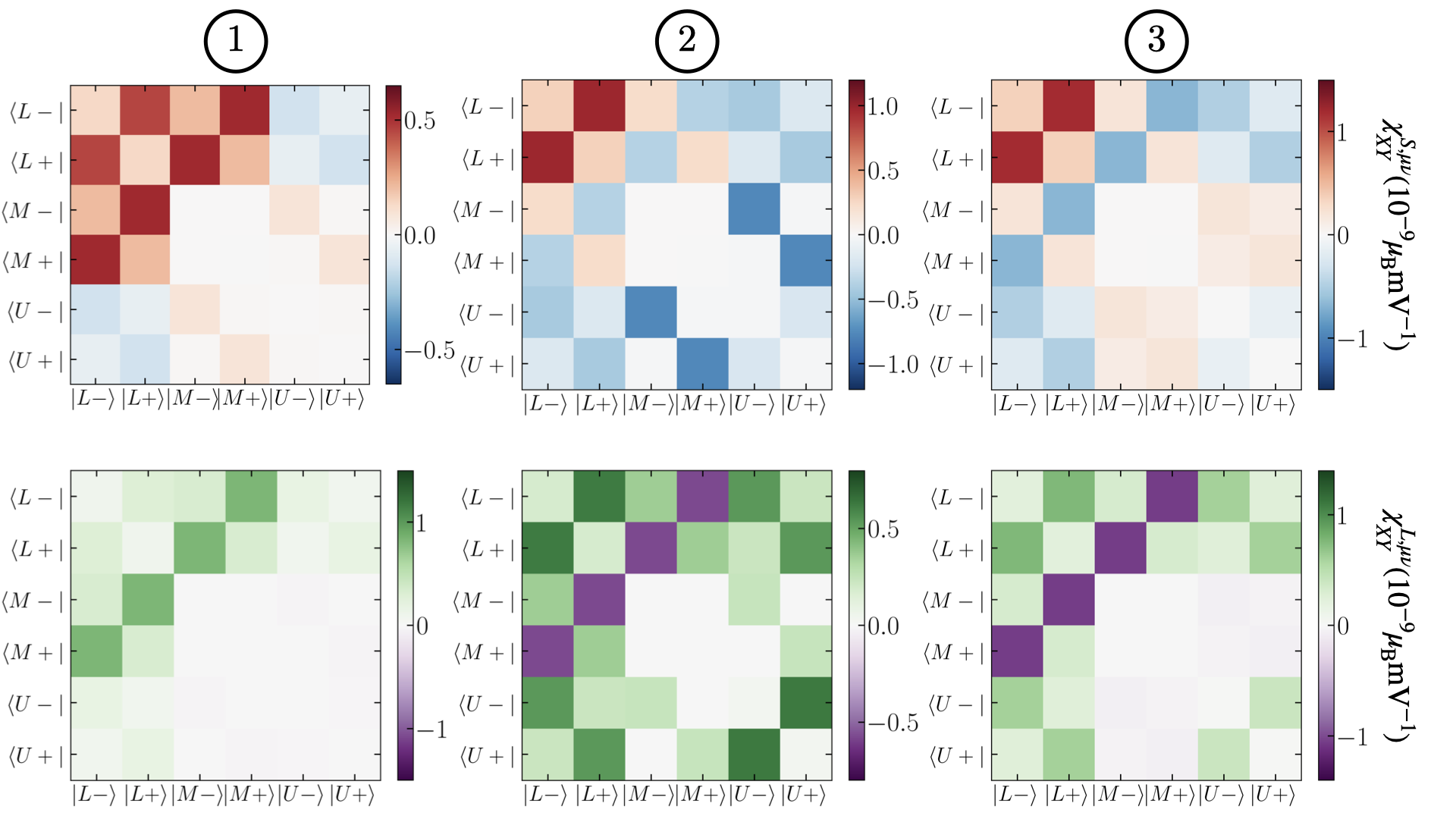}
    \caption{Spin (upper panel) and orbital (lower panel) Edelstein susceptibility projected over the $L$, $M$ and $U$ states. The chemical potential $\mu$ is fixed at values 1,2 and 3 referring to Fig.~\ref{band_structure}}
    \label{fig_mat}
\end{figure*}
\\These textures are responsible for the spin and orbital EE, when an electric field is included into the system. In linear response theory the magnetization $m_\alpha$ along the $\alpha$ direction is
\begin{equation}
    m_\alpha=\chi_{\alpha\beta}E_{\beta},
\end{equation}
where $\chi_{\alpha\beta}$ is the Edelstein susceptibility and $E_{\beta}$ is the electric field in the $\beta$ direction. $\chi_{\alpha\beta}$ is the sum of two contributions: a spin contribution $\chi_{\alpha\beta}^S$ and an orbital one $\chi_{\alpha\beta}^L$.
We use the Boltzmann approach within the time relaxation approximation to compute the Edelstein susceptibility~\cite{Bibes_edelstein}. 
The magnetic moment per unit cell in the crystal is
\begin{equation}
    m_\alpha=\frac{\mu_b}{\hbar}S_{cell}\sum_{n}\int_{BZ}\frac{d^2\vek}{(2\pi\tilde{a})^2}\hspace{0.2cm}\delta f(\vek)\langle 2S_\alpha+L_{\alpha}\rangle_{n}(\vek)
\end{equation}
where $\langle S_i\rangle_{n}(\vek)$ is the mean value over the eigenstates of the $n$-th band, $\mu_b$ is the Bohr magneton, $\hbar$ is the reduced Plank's constant, $S_{cell}$ is the unit cell area, and $\delta f$ is the modification of the thermal distribution $f_{\rm{th}}$ in linear response regime, which is expressed as
\begin{equation}
    \delta f(\vek)=-\tau_0 q\tilde{a} \Vec{E}\cdot\frac{\partial f_{\text{th}}}{\partial\hbar\Vec{k}}.
\end{equation}
Here $\tau_0$ is the relaxation time, $q$ is the charge of the electrons.
Therefore
\begin{equation}
    \chi_{\alpha\beta}^{\mathcal{O}}=\left(-\frac{\tau_0 q\mu_b}{\tilde{a}\hbar^2}S_{cell}\sum_{n}\int_{BZ}\frac{d^2\vek}{(2\pi)^2}\hspace{0.2cm}\frac{\partial f_{\text{th}}}{\partial k_\beta}\langle \mathcal{O}_\alpha\rangle_{n}(\vek)\right),
    \label{chi}
\end{equation}
where $\mathcal{O}_\alpha=2S_\alpha$ or $L_\alpha$. Due to the anti-symmetric property of the $\chi_{\alpha\beta}$~\cite{shen2014}, we need only to evaluate $\chi_{XY}$ (with $X=(\bar{1}10)$ and $Y=(\bar{1}\bar{1}2$) directions).
The results are collected in Fig.~\ref{fig:Edelstein_chi} both for the spin and the orbital susceptibility as a function of the chemical potential $\mu$. We fixed the temperature to $T=10$ K and $\tau_0=3.4\times10^{-12}$~s, value which is derived from the experimental mobility in Ref.~\cite{khan2017high}.
Both susceptibilities behave non monotonically and they are explicitly decomposed into the contributions of the three Kramers doublets in Eq.~(\ref{chi}), as also done in \cite{Bibes_edelstein}. 
The spin susceptibility changes sign and presents a maximum and a minimum suggesting that, in real systems, a magnetization reversal can be induced by appropriate gating (e.g. back gate control of the chemical potential).
On the other hand the orbital susceptibility is always of the order of $10^{-8}\mu_B$ mV$^{-1}$, which is one order of magnitude greater than the spin susceptibility above the benchmark line $3$ of Fig.~\ref{fig:Edelstein_chi}.
\\We demonstrate that a crucial ingredient for our results is the intermixing between different doublets. 
%In the Supplementary material~\cite{supplementary} we show an analysis for a single Rashba doublet. We demonstrate that even taking three separate doublets can not reproduce the full behaviour of the conductance~\ref{chi}. 
The reason is that the orbital Rashba term $\hat{L}\times\vek$ induces an orbital angular momentum which is larger where the doublets are maximally hybridized.
To demonstrate the role of the hybridization, we introduce $P^{\mu}=\ket{\mu}\bra{\mu}$ projector along the eigenstate $\mu$ of Hamiltonian~(\ref{hamiltonian_comp}) evaluated for $\vek=0$. In this case, we identify three different states twice degenerate that we call $\ket{L\sigma}$ (as low), $\ket{M\sigma}$ (as middle), and $\ket{U\sigma}$ (as up). We decompose the spin operator $S$ as
\begin{equation}
    S_\alpha=\sum_{\mu\nu}P^{\mu}S_\alpha P^{\nu}=\sum_{ij}S_\alpha^{\mu\nu}.
    \label{Spin_proj}
\end{equation}
A similar decomposition is adopted for L.
By substituting Eq.~(\ref{Spin_proj}) in Eq.~(\ref{chi}) one can define an Edelstein susceptibility projected on the states for $\vek=0$,  $\chi_{\alpha\beta}^{\mathcal{O},\mu\nu}$, respecting the condition
\begin{equation}
    \chi^{\mathcal{O}}_{\alpha\beta}=\sum_{\mu\nu}\chi_{\alpha\beta}^{\mathcal{O},\mu\nu}.
\end{equation}
The magnitude of this quantity is an indicator of how much the hybridization of the doublets is important for EE or OEE.
%This response is an indication of how much the \MT{hybridization} contributes to the \MT{spin and orbital} magnetization.
The values of $\chi_{XY}^{\mathcal{O},\mu\nu}$ for different benchmark chemical potential are represented in Fig.~\ref{fig_mat}.
Thus we conclude that for the first red peak of Fig.~\ref{fig:Edelstein_chi}(a) there is a strong connection between the first two bands, indicating that this peak is described by a single doublet. However there is the presence of hybridization with the second doublet too, which is of the same order of magnitude of the intra-doublet interaction. The effect is even more evident for the angular momentum. By increasing the chemical potential, more doublets are filled and the hybridization becomes more relevant. However it is always true that the second intra-doublet contribution is zero, as seen from the $2\times2$ white square in Fig.~\ref{fig_mat}. This is a direct consequence of the absence of linear and quadratic splitting for the two bands in the second doublet. The intermediate doublet mediates the interaction between the first and the third doublet. This is confirmation of the relevance of the multiband model. Differently from the (001) interface, the (111) interface has the three doublets relatively close one to each other, leading to this strong orbital hybridization.

\noindent\textit{Discussion.} We have shown that the multiband character of the (111) LAO/STO is a key-feature for the emerging non-linear spin and orbital EE. The strong SOC and the confining potential lead to a non-trivial Rashba interaction. Together with the orbital hybridization of the bands, this allows a spin and orbital magnetic moment in the presence of an in-plane external magnetic field. 
We have shown through a tight-binding model that the generalized Rashba effect can generate a non-linear spin EE which changes its sign with the chemical potential, and that can be modulated with an external gate. Moreover, the strong orbital character of the bands leads to an OEE, an order of magnitude higher than the spin effect. 
Up to now there is no direct evidence of orbital magnetization in the experiments. In fact, since one can observe only the full magnetization it is difficult to disentangle the spin from the orbital response~\cite{revieworbitronic2021}. However, by tuning to zero the spin Edelstein susceptibility one can disentangle the two components, overcoming this problem. In the supplementary material~\cite{supplementary} we show how a $\vek$-dependency on the scattering time changes the results. In principle, the strong orbital degeneracy and the hybridization of the bands could be enhanced by taking into account the contribution by impurities. However, direct computation within the framework of a $k$-dependent relaxation time shows small quantitative modification of results presented in the main text of this work. In particular $\chi^S_{XY}$ vanishes at the same energy values predicted within the framework of a constant $\tau$ theory. Thus, our results provide a consistent picture of the system response. The proposal of tuning the spin response to zero, together with the new proposal of measuring the so-called orbital torque \cite{bhowal2021orbital}, makes the (111) LAO/STO interface suitable for investigating the orbital magnetization and represents a promising spin-orbitronic platform. We remark that (111) KaTiO$_3$-based heterostructures~\cite{bareille2014two,bruno2019band,vicente2021spin} have a similar crystalline structure with higher SOC, which could enhance the EE and OEE. Therefore, they could be an interesting system where further apply our analysis.

\appendix
\clearpage
\onecolumngrid
\section{Details of the model}
The effective 2D single-particle Hamiltonian originating from the three $t_{2g}$ orbitals of the Ti-atoms in the bilayer reads~\cite{xiao2011interface,trama2021straininduced}
\begin{equation}
    H=H_{\text{TB}}(t_D,t_I)+H_{\text{SOC}}(\lambda)+H_{\text{TRI}}(\Delta)+H_{v}(v),
\end{equation}
%for which we have highlighted the dependence on the parameters for ease of discussion.
where $H_{\rm{TB}}$ is the hopping Hamiltonian which in $\vek$-space can be written as:
\begin{equation}
    H_{\text{TB}}=\sum_{\vek}\sum_{i,\alpha\beta,\sigma}t_i^{\alpha\beta}(t_D,t_I,\Vec{k}) d_{i\alpha\sigma,\vek}^{\dagger} d_{i\beta\sigma,\vek}
    \label{TBk},
\end{equation}
where $d_{i\alpha \sigma,\vek}$ is 
the annihilation operator of
the electron with 2D dimensionless quasi-momentum $\vek=\Tilde{a}\Vec{K}$,
where $\Vec{K}$  is the quasi-momentum, occupying the orbital
$i = {xy, yz, zx}$ belonging to the layer $\alpha,\beta= {\text{Ti}_1, \text{Ti}_2}$ and of spin $\sigma= \pm 1/2$. The matrix $t_i^{\alpha\beta}(t_D,t_I,\Vec{k})$, in the basis $\{d_{yz},d_{zx},d_{xy}\}\otimes\{\text{Ti}_1,\text{Ti}_2\}$ is the following
\begin{equation}
    t_i^{\alpha\beta}=\begin{pmatrix}
        0 & 0 & 0 & \epsilon_{yz} & 0 & 0 \\
        0 & 0 & 0 & 0 & \epsilon_{zx} & 0 \\
        0 & 0 &0 & 0 & 0 & \epsilon_{xy} \\
        \epsilon_{yz}^* & 0 & 0 & 0 & 0 & 0 \\
        0 & \epsilon_{zx}^* & 0 & 0 & 0 & 0 \\
        0 & 0 & \epsilon_{xy}^* & 0 & 0 & 0 \\
    \end{pmatrix},
\end{equation}
where the interlayer contributions are:
\begin{eqnarray}
     \nonumber &\epsilon_{yz}&=-t_D\left(1+e^{i(\frac{\sqrt{3}}{2}k_X-\frac{3}{2}k_Y)}\right)-t_Ie^{-i(\frac{\sqrt{3}}{2}k_X+\frac{3}{2}k_Y)},\\
     &\epsilon_{zx}&=-t_D\left(1+e^{-i(\frac{\sqrt{3}}{2}k_X+\frac{3}{2}k_Y)}\right)-t_Ie^{i(\frac{\sqrt{3}}{2}k_X-\frac{3}{2}k_Y)},\\
     \nonumber &\epsilon_{xy}&=-2t_D\cos(\frac{\sqrt{3}}{2}k_X)e^{-i\frac{3}{2}k_Y}-t_I.
     \label{interlayer}
\end{eqnarray}
The direct $t_D$ and indirect $t_I$ couplings have been fixed to the values $t_D=0.5$ eV and $t_I=0.04$ eV~\cite{trama2021straininduced} via comparison with angular resolved photoemission spectroscopy data.
\\$H_{\rm{SO}}$ is the atomic SOC coupling, which has the following expression
\begin{equation}
    H_{\text{SOC}}=\frac{\lambda}{2}\sum_{\vek}\sum_{ijk,\alpha,\sigma\sigma'}i\varepsilon_{ijk}
    d_{i\alpha\sigma,\vek}^{\dagger} \sigma^{k}_{\sigma\sigma'}d_{j\alpha\sigma',\vek}
    \label{eq:spinorbit}
\end{equation}
where $\varepsilon_{ijk}$ is the Levi-Civita tensor, and $\sigma^k$ are the Pauli matrices. We fix the SOC coupling $\lambda=0.01$ eV, as a typical order of magnitude~\cite{monteiro2019band}.
\\The trigonal crystal field Hamiltonian $H_{\rm{TRI}}$ takes into account the strain at the interface along the (111) direction. The physical origin of this strain is the possible contraction or dilatation of the crystalline planes along the (111) direction. This coupling has the form~\cite{khomskii2014transition}
\begin{equation}
    H_{\text{TRI}}=\frac{\Delta}{2}\sum_{\vek}\sum_{i\neq j,\alpha,\sigma} d_{i\alpha\sigma,\vek}^{\dagger} d_{j\alpha\sigma,\vek}.
    \label{eq:trigonal}
\end{equation}
We fix $\Delta=-0.005$ eV as reported in~\cite{de2018symmetry}.
\\Finally the last term $H_v$ describes an electric field in the (111) direction, orthogonal to the interface,
which breaks the reflection symmetry.
%Differently from the previous contributions %\cite{xiao2011interface,monteiro2019band}, we treat the effect of %the electric field on the orbitals perturbatively. In particular, 
The Hamiltonian $H_v$ can thus be written as the sum of an electrostatic potential $H_{v0}$ and a term which induces the breaking of the inversion symmetry in the orbitals $H_{\text{BIS}}$
\begin{equation}
    H_v=\frac{v}{2}\sum_{i,\alpha,\sigma,\vek}\xi_{\alpha} d_{i\alpha\sigma,\vek}^{\dagger} d_{i\alpha\sigma,\vek}+
   \sum_{\vek}\sum_{ij,\alpha\beta,\sigma}h_{ij,\vek}^{\alpha\beta}(v) d_{i\alpha\sigma,\vek}^{\dagger} d_{j\beta\sigma,\vek}=H_{v0}+H_{\text{BIS}},
    \label{electric}
\end{equation}
where $\xi_{\text{Ti}_1/\text{Ti}_2}=\pm 1$. For ease of writing $h_{ij,\vek}^{\alpha\beta}(v)$ is written as the sum of two components: an interlayer contribution, connecting the two layers Ti$_1$ and Ti$_2$, as
\begin{equation}
   h_{ij,\vek}^{\text{Ti}_1\text{Ti}_2}=h_{ij,\vek}^{\text{Ti}_2\text{Ti}_1}=
    \eta_p\frac{V_{pd\pi}(\sqrt{2})^{7/4}}{\sqrt{15}}
    \begin{pmatrix}
    0 & -2i e^{i\frac{3}{2}k_Y} \sin{(\frac{\sqrt{3}}{2}k_X)}& 1-e^{\frac{i}{2}\left(\sqrt{3}k_X+3k_Y\right)} \\
    2i e^{i3/2k_Y} \sin{(\frac{\sqrt{3}}{2}k_X)} & 0 & 1-e^{-\frac{i}{2}\left(\sqrt{3}k_X-3k_Y\right)}\\
    -1+e^{\frac{i}{2}\left(\sqrt{3}k_X+3k_Y\right)}  & -1+e^{-\frac{i}{2}\left(\sqrt{3}k_X-3k_Y\right)} & 0\\
    \end{pmatrix},
    \label{t1hopt2}
\end{equation}
and an interlayer contribution 
as $h_{ij,\vek}^{\text{Ti}_1\text{Ti}_1}=h_{ij,\vek}^{\text{Ti}_2\text{Ti}_2}=h_{ij,\vek}^\pi+h_{ij,\vek}^\sigma$ where
\begin{equation}\small
    h_{ij,\vek}^{\pi}=\eta_p\frac{2i}{\sqrt{15}}V_{pd\pi}
    \begin{pmatrix}
        0 & -(\sin(\kappa_1)+\sin(\kappa_2)+2\sin(\kappa_3)) & (\sin(\kappa_1)+2\sin(\kappa_2)+\sin(\kappa_3))\\
        (\sin(\kappa_1)+\sin(\kappa_2)+2\sin(\kappa_3)) & 0 & -(2\sin(\kappa_1)+\sin(\kappa_2)+\sin(\kappa_3))\\
        (\sin(\kappa_1)+2\sin(\kappa_2)+\sin(\kappa_3)) & (2\sin(\kappa_1)+\sin(\kappa_2)+\sin(\kappa_3)) & 0
        \end{pmatrix},\normalsize
    \label{electric_pi}
\end{equation}
\begin{equation}
    h_{ij,\vek}^{\sigma}=\eta_p\frac{2i}{\sqrt{15}}\sqrt{3}V_{pd\sigma}
    \begin{pmatrix}
        0 & (\sin(\kappa_1)+\sin(\kappa_2)) & -(\sin(\kappa_1)+\sin(\kappa_3))\\
        -(\sin(\kappa_1)+\sin(\kappa_2)) & 0 & (\sin(\kappa_2)+\sin(\kappa_3))\\
        (\sin(\kappa_1)+\sin(\kappa_3)) & -(\sin(\kappa_2)+\sin(\kappa_3)) & 0
        \end{pmatrix},
    \label{popo}
\end{equation}
with
$\kappa_1=-\frac{\sqrt{3}}{2}k_X+\frac{3}{2}k_Y$, $\kappa_2=-\frac{\sqrt{3}}{2}k_X-\frac{3}{2}k_Y$, $\kappa_3=\sqrt{3}k_X$, $V_{pd\pi}=0.028$ eV and $V_{pd\sigma}=-0.065$ eV, $\eta_p\sim\frac{v\sqrt{3}}{a_0} \frac{1}{10\text{ eV/nm}}\sim0.09$ by using $v=0.2$ eV and $a_0=3.905$ nm.
The electric field has been fixed at the value $v=0.2$ eV by comparison with the Rashba splitting evaluated in Ref.~\cite{monteiro2019band}. The full derivation of Eq.~(\ref{electric}) can be found in the Supplementary material of Ref.~\cite{trama2022gate}.
\subsection*{Expansion at low fillings}
The whole matrix $H_{\text{TB}}+H_{v0}$, which is $12\times12$, admits as eigenstates
\begin{equation}
    \ket{\psi_{i\sigma,\vek}{\pm}}=\alpha_i(\vek)e^{i\phi_i(\Vec{k})}\ket{d_{i1\sigma,k}}+ \beta_i^{\pm}(\vek) \ket{d_{i2\sigma,k}}
    \label{vectors66},
\end{equation}
with
\begin{align}\label{coefficients66}
      \nonumber&\alpha_i^{\pm}(\vek)=\frac{|\epsilon_i(\vek)|}{\sqrt{2|\epsilon_i(\vek)|^2+\frac{v^2}{2}\pm v\sqrt{\frac{v^2}{4}+|\epsilon_i(\vek)|^2}}};
      \\
      &\beta_i^{\pm}(\vek)=\frac{\left(\frac{v}{2}\pm\sqrt{\frac{v^2}{4}+|\epsilon_i(\vek)|^2}\right)}{\sqrt{2|\epsilon_i(\vek)|^2+\frac{v^2}{2}\pm v\sqrt{\frac{v^2}{4}+|\epsilon_i(\vek)|^2}}};
      \\
      \nonumber&\phi_i(\Vec{k})=\arg[\epsilon_i(\vek)]
\end{align}
where the orbitals are labeled by the index $i$ and the spin using the index $\sigma$.
The corresponding eigenvalues are
\begin{equation}
    \rho_i^{\pm}(\vek)=\pm\sqrt{\frac{v^2}{4}+|\epsilon_i(\vek)|^2},
    \label{eigenval66}
\end{equation}
which expanded at the second order in $\vek$ gives the expression $\mathcal{E}_i$ of Eqs.~(\ref{secondorderTB1}-\ref{secondorderTB3}) in the main text.
\\In order to obtain the orbital Rashba Hamiltonian on the six lower bands for low fillings, we simultaneously linearize the Electric field Hamiltonian $H_{\text{BIS}}$ of Eq.~(\ref{electric}) as a function of $\vek$ and evaluate its matrix elements among the six lower states in Eq.~(\ref{vectors66}), evaluated for $\vek=0$. The result is the following linear Hamiltonian:
\begin{equation}
\centering
    (H_{\text{BIS}})_{ij}=-i\mathcal{F}\varepsilon_{ijk}\kappa_k, \quad \text{where}\quad
    \mathcal{F}=\frac{2\eta_p}{\sqrt{15}}\left(V_{pd\pi}(1+2^{7/8}\alpha\beta\cos{(\phi}))+\sqrt{3}V_{pd\sigma}\right),
\end{equation}
$\Vec{\kappa}=(\kappa_1,\kappa_2,\kappa_3)$ as defined above and $\alpha$, $\beta$ and $\phi$ are the Eqs.~(\ref{coefficients66}) evaluated for $\vek=0$.
\\Identifying now the matrix elements of the orbital angular momentum $\hat{L}$ matrices, we can rewrite this term as:
\begin{equation}
    H_{\text{BIS}}=\frac{3}{\sqrt{2}}\mathcal{F} (\vek\times\hat{L})\cdot\hat{n}_{111},
    \label{eq_kl}
\end{equation}
where $\hat{n}_{111}$ is a unitary vector along the $(111)$ direction.
\\Having introduced the notation of the angular momentum we can write also the TB over the states~(\ref{vectors66}) using the same notation:
\begin{equation}
    H_{\text{TB}}=\sum_{i}\mathcal{E}_i (\mathbb{1}-L_i^2)\otimes\mathbb{1}_{\sigma\sigma'}.
    \label{TBintermsofL}
\end{equation}
Also $H_{\text{TRI}}$ can be expressed in the form
\begin{equation}
    H_{\text{TRI}}=\Delta(\scalebox{1.25}{$\mathbb{1}$}-\frac{3}{2}L_{111}^2).
    \label{eq_app_delta}
\end{equation}

\section{Spin and orbital textures}
\begin{figure*}
    \centering
    \includegraphics[width=0.85\textwidth]{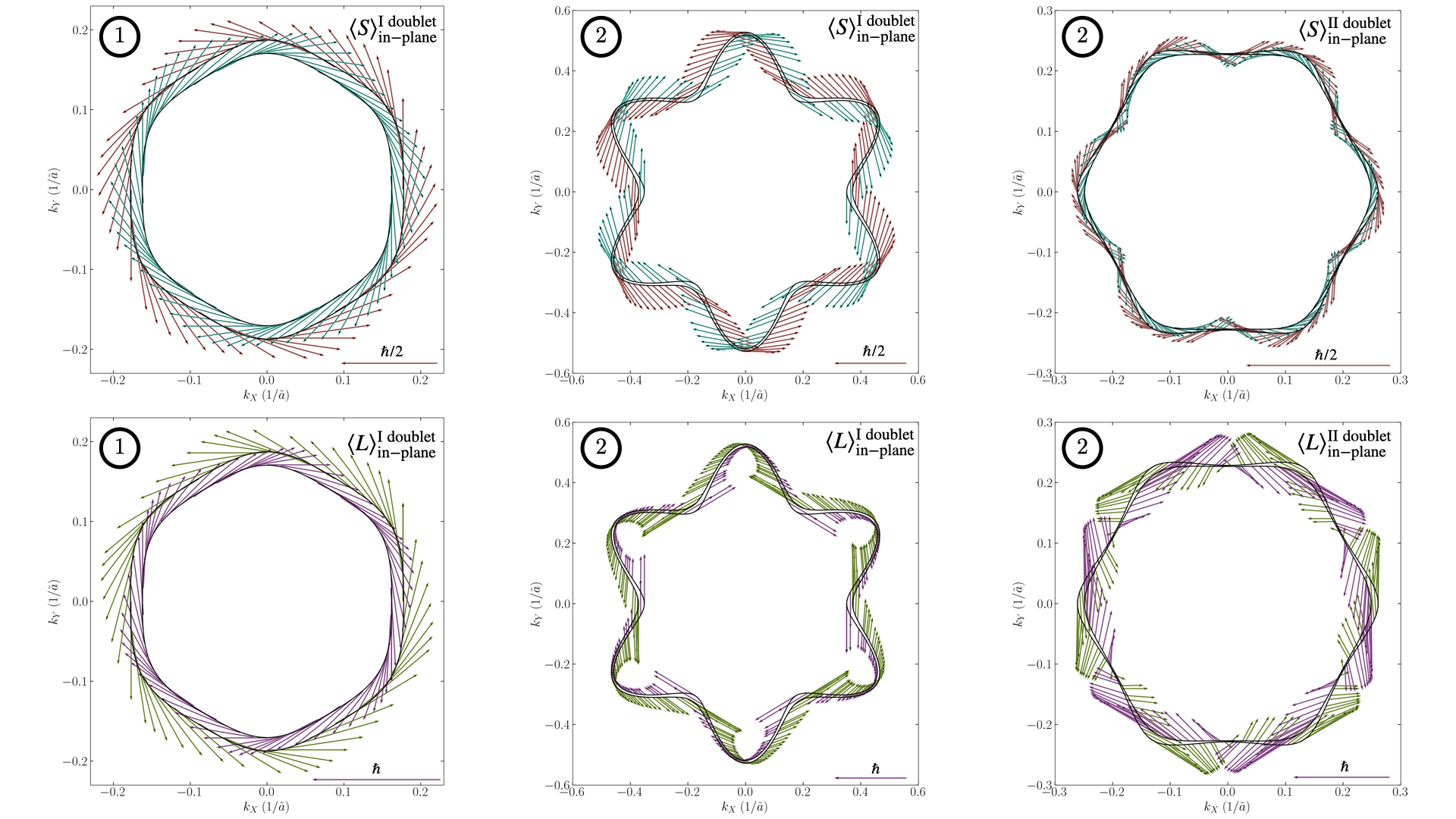}
    \caption{In-plane spin (Upper panel) and orbital angular momentum (Lower panel) textures for the first and the second doublet with the chemical potential fixed to the value corresponding to the benchmark line 1 and 2 in Fig.~\ref{band_structure}. The red and green arrows represent the mean value of in-plane component of the operator for the external band, while the blue and pink refer to the internal component. The mean value of the generic operator $O$ is evaluated as $\langle O\rangle=\sqrt{\langle O_{\overline{1}10}\rangle^2+\langle O_{\overline{11}2}\rangle^2}$.}
    \label{Pattern_altri}
\end{figure*}

In Fig.~\ref{Pattern_altri} we report the spin and orbital angular momentum textures for the benchmark lines 1 and 2 of the Fig.~\ref{band_structure}. For the benchmark line 1 the spin and the angular momentum follows the same pattern, that reflects the dominance of the atomic SOC in this region: the total angular momentum $\hat{J}$ is the conserved quantity and therefore both $\langle\hat{L}\rangle$ and $\langle\hat{S}\rangle$ are proportional to $\langle\hat{J}\rangle$. For the benchmark line 2, a second doublet is occupied and the hybridization of the bands is present. In this case the patterns for the spin and the orbital angular momentum are different. The in-plane $\langle \hat{L}\rangle$ is higher where the bands are maximally hybridized.

\section{Role of impurities scattering on the Edelstein effect}
\begin{figure}[b]
	\includegraphics[width=0.68\textwidth]{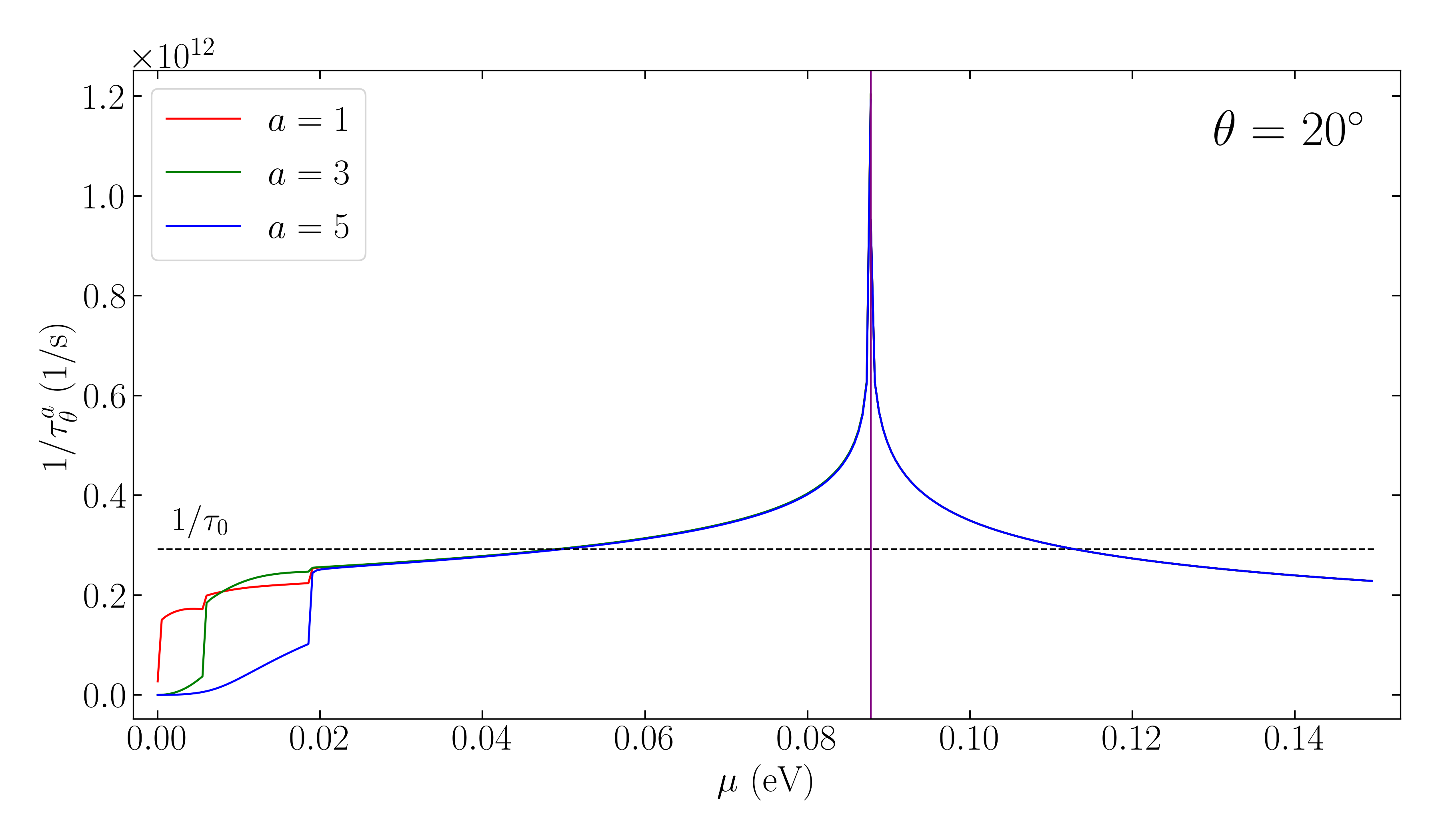}
	\caption{$1/\tau$ as a function of the chemical potential for a benchmark direction in the Brillouin zone. The dashed line corresponds to the inverse of the scattering time used in the main text. The purple vertical line corresponds to the energy at which a Lifshitz transition occurs for first band (see Fig.~\ref{point_like_edelst}).}
	\label{scattering_point}
\end{figure}
\begin{figure}
    \centering
    \includegraphics[width=0.85\textwidth]{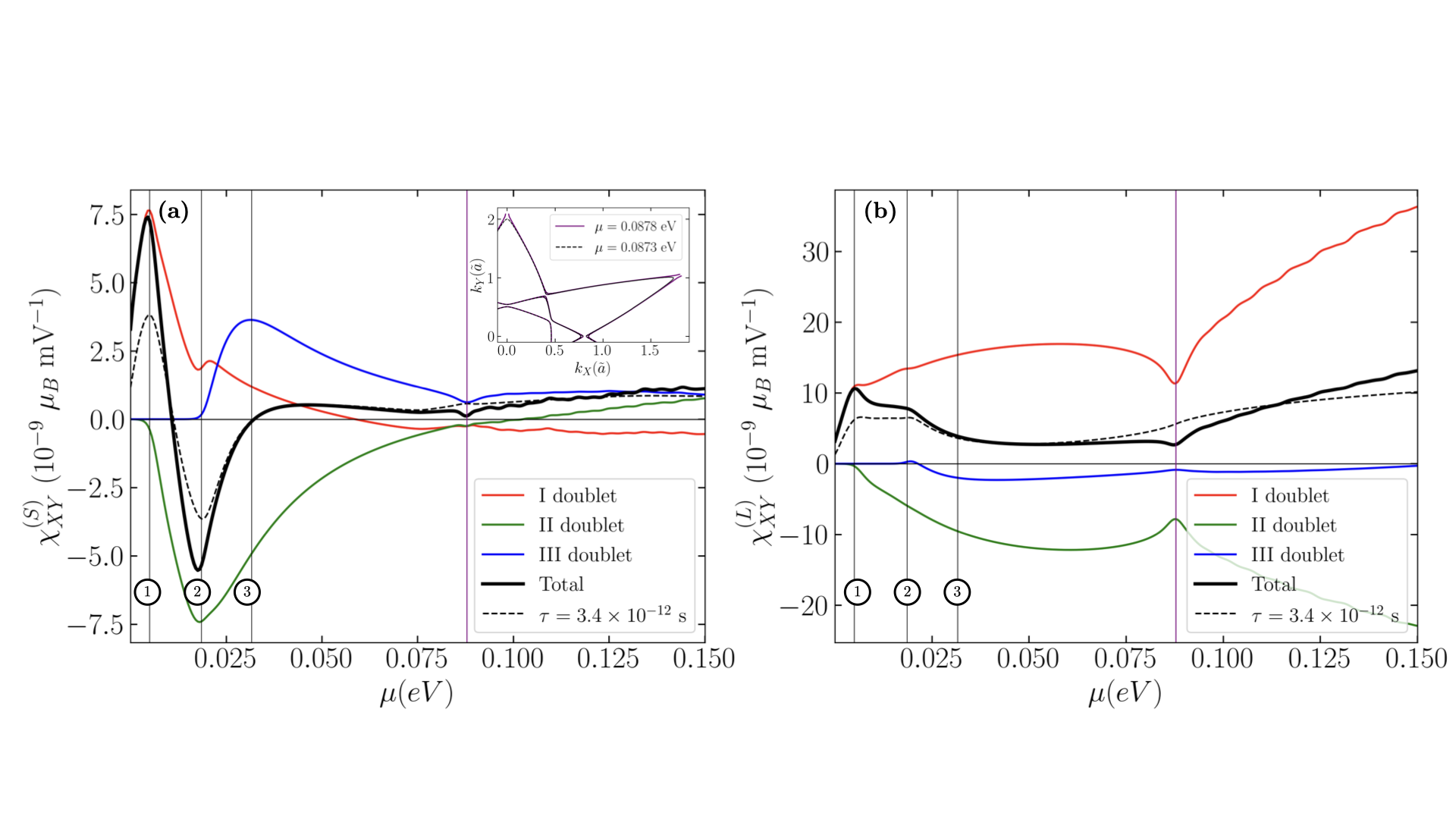}
    \caption{Spin (a) and orbital (b) Edelstein coefficient  as a function of the chemical potential using a scattering time model with a point-like impurity. The different colors correspond to the contribution of a specific Kramers doublet, while the dashed line corresponds to the total Edelstein response for a constant $\tau$. The inset in panel (a) corresponds to the detail of the Fermi energy contour for the purple chemical potential line.}
    \label{point_like_edelst}
\end{figure}

In the main text we assumed $\tau$ to be independent of the quasi-momentum, following the assumptions of the previous work in Ref.~\cite{Bibes_edelstein}. In this section we show the prediction for the Edelstein susceptibility when the dependence of the scattering time $\tau$ on $\vek$ is included. We consider a point-like impurity, e.g. oxygen vacancy, a kind of defect typical of SrTiO$_3$ heterostructures~\cite{kalabukhov2007effect,li2011formation,walker2014control}.
\\The unperturbed Hamiltonian is
\begin{equation}
    H_\vek=\sum_{\alpha\beta\vek}h_{\alpha\beta\vek}\cdag{\alpha\vek}c_{\beta\vek}=\sum_{a\vek}\epsilon_{a\vek}\cdag{a\vek}c_{a\vek},
\end{equation}
where the Greek indices label the orbital degrees of freedom, while the Latin ones label the diagonalized bands. The change of basis due to the diagonalization is ruled by the following transformation
\begin{equation}
    c_{a\vek}=U^\vek_{a,\alpha}c_{\alpha\vek}.
    \label{U_bas}
\end{equation}
An impurity can occupy the position of an atom in the lattice. Since the elementary cell contains two atoms (Ti$_1$ or Ti$_2$) we will distinguish the degrees of freedom as $\alpha=o,t$, where $t$ labels the kind of atom in the unit cell, while $o$ labels all the other degrees of freedom (spin and orbitals).
\\With the previous definitions, we write the impurity Hamiltonian of $N$ impurities which can randomly occupy an atom position as
\begin{equation}
    H_{\rm{I}}=\sum_{i,oo^\prime}\frac{\epsilon^I_{oo^\prime}}{v} \cdag{ot_i,r_i}c_{o^\prime t_i,r_i}=
    \sum_{\vek\;\vec{q}}\sum_{i,oo^\prime}\frac{\epsilon^{I}_{oo^\prime}}{v}\cdag{o t_i\vek}c_{o^\prime t_i\vec{q}}e^{i\frac{(\vec{q}-\vek)}{\tilde{a}}\cdot \vec{r}_i}=
    \sum_{\vek\; \vec{q}}\sum_{i,a b} \frac{\varepsilon_{ab,t_i}^{\vek \vec{q}}}{v}\cdag{a\vek}c_{b\vec{q}}e^{i\frac{(\vec{q}-\vek)}{\tilde{a}}\cdot \vec{r}_i},
\end{equation}
where we summed over all the impurities $i$ in the first sum, defined $v$ the volume of the crystal, $\epsilon^I_{oo^\prime}=\int d^3\vec{r}\;V_{oo^\prime}(\vec{r}-\vec{r}_i)$, $V_{oo^\prime}(\vec{r}-\vec{r}_i)$ the impurity potential, $\vec{r}_i$ the position of the impurity, and $\varepsilon_{ab,t_i}^{\vek \vec{q}}=\sum_{oo^\prime}\epsilon_{oo^\prime}^I U_{a,ot_i}^\vek U_{b,o^\prime t_i}^{\vec{p}\dagger}$. 
We do not sum over the label $t_i$ because every impurity occupies only one layer.
We take the following form for the impurity energy tensor 
\begin{equation}
	\epsilon^I_{oo^\prime}=\epsilon_0\delta_{oo^\prime}.
\end{equation}
This expression is the simplest coupling for the scattering: the electron maintains its spin and orbital character, preserving the symmetries of the system. This choice allows us to have control on the results and see the differences with respect to the constant $\tau$.
With such a definition we can perform a diagrammatic calculation for evaluating the self-energy due to the impurities. We perform an average over all the positions $r_i$ and all the layer occupation $t_i$, which are independent distributions.
The imaginary part of the self-energy for the $a$-th band located at the $t$ layer, up to the second order of perturbation theory over the strength of the impurity potential, is
\begin{equation}
    \text{Im}({\Sigma^a_{t,\vek}})=-\frac{\hbar}{2\tau^a_{\vek}}=-n_i\frac{\varepsilon_0^2}{v}\pi\sum_{\vec{p}}\sum_b\delta(\hbar\omega_{\vek}^a-\hbar\omega_{\vec{p}}^b)|\sum_{oo^\prime} U_{a,ot}^\vek U_{b,o t}^{\vec{p}\dagger}|^2,
\end{equation}
where $n_i$ is the impurity density.
This coincides with the inverse of the scattering time, and is shown in Fig.~\ref{scattering_point} as a function of the chemical potential $\mu$ (or $\hbar\omega_{\vek}^a$) for some benchmark directions in the BZ.
The averaged $1/\tau_\vek^a$ over the $t$ layers is therefore
\begin{equation}
    \frac{1}{\tau_\vek^a}=\left( \frac{1}{\tau_{\text{Ti}_1\vek}^a}+ \frac{1}{\tau_{\text{Ti}_2\vek}^a}\right)\frac{1}{2}.
\end{equation}
The magnitude of the coupling $\epsilon_0$ is determined by averaging $\tau_\vek$ over the whole BZ, and fixing the mean value at the experimental value $\tau_0$~\cite{khan2017high} used in the main text.
With such a scattering time the Edelstein response is
\begin{equation}
    \chi_{\alpha\beta}^{\mathcal{O}}=-\frac{ q\mu_b}{\tilde{a}\hbar^2}S_{cell}\sum_{n}\int_{BZ}\frac{d^2\vek}{(2\pi)^2}\hspace{0.2cm}\frac{\partial f_{th}}{\partial k_\beta}\tau_n(\vek)\langle \mathcal{O}_\alpha\rangle_{n}(\vek).
    \label{chi_tau_k}
\end{equation}
The results obtained using a temperature of $T=10$ K are showed in Fig.~\ref{point_like_edelst}.
Here we can see that impurities do not substantially modify the behaviour of the curves, but change the response quantitatively, especially for low fillings. 
We notice the appearance of a local minima at the second benchmark chemical potential for the first doublet and a drop of the response at the purple vertical line. In both cases the susceptibility has a drop due to the large number of states available for the scattering, which reduce the scattering lifetime and consequently the Edelstein response. In the first case, the maximal mixing between the bands induces the appearance of the local minima, as shown in Fig.~\ref{fig_mat}, while in the second case, the drop is due to the Lifshtiz transition for the first doublet as shown in Fig~\ref{point_like_edelst}. 

\bibliographystyle{unsrt}
\bibliography{Bib}

\begin{thebibliography}{10}

\bibitem{dieny2020opportunities}
Bernard Dieny, Ioan~Lucian Prejbeanu, Kevin Garello, Pietro Gambardella, Paulo
  Freitas, Ronald Lehndorff, Wolfgang Raberg, Ursula Ebels, Sergej~O
  Demokritov, Johan Akerman, et~al.
\newblock Opportunities and challenges for spintronics in the microelectronics
  industry.
\newblock {\em Nature Electronics}, 3(8):446--459, 2020.

\bibitem{wolf2001spintronics}
SA~Wolf, DD~Awschalom, RA~Buhrman, JM~Daughton, von~S von Moln{\'a}r,
  ML~Roukes, A~Yu Chtchelkanova, and DM~Treger.
\newblock Spintronics: a spin-based electronics vision for the future.
\newblock {\em science}, 294(5546):1488--1495, 2001.

\bibitem{dyakonov1971current}
Mikhail~I Dyakonov and VI~Perel.
\newblock Current-induced spin orientation of electrons in semiconductors.
\newblock {\em Physics Letters A}, 35(6):459--460, 1971.

\bibitem{edelstein1990spin}
Victor~M Edelstein.
\newblock Spin polarization of conduction electrons induced by electric current
  in two-dimensional asymmetric electron systems.
\newblock {\em Solid State Communications}, 73(3):233--235, 1990.

\bibitem{aronov1989nuclear}
AG~Aronov and Yu~B Lyanda-Geller.
\newblock Nuclear electric resonance and orientation of carrier spins by an
  electric field.
\newblock {\em Soviet Journal of Experimental and Theoretical Physics Letters},
  50:431, 1989.

\bibitem{supplementary}
M.~Trama and al.
\newblock Supplemental material.

\bibitem{lin2019interface}
Weinan Lin, Lei Li, Fatih Do{\u{g}}an, Changjian Li, H{\'e}l{\`e}ne Rotella,
  Xiaojiang Yu, Bangmin Zhang, Yangyang Li, Wen~Siang Lew, Shijie Wang, et~al.
\newblock Interface-based tuning of rashba spin-orbit interaction in asymmetric
  oxide heterostructures with 3d electrons.
\newblock {\em Nature communications}, 10(1):1--7, 2019.

\bibitem{hwang2012emergent}
Harold~Y Hwang, Yoh Iwasa, Masashi Kawasaki, Bernhard Keimer, Naoto Nagaosa,
  and Yoshinori Tokura.
\newblock Emergent phenomena at oxide interfaces.
\newblock {\em Nature materials}, 11(2):103--113, 2012.

\bibitem{caviglia2010tunable}
AD~Caviglia, M~Gabay, Stefano Gariglio, Nicolas Reyren, Claudia Cancellieri,
  and J-M Triscone.
\newblock Tunable rashba spin-orbit interaction at oxide interfaces.
\newblock {\em Physical review letters}, 104(12):126803, 2010.

\bibitem{gariglio2018spin}
Stefano Gariglio, AD~Caviglia, Jean-Marc Triscone, and M~Gabay.
\newblock A spin--orbit playground: Surfaces and interfaces of transition metal
  oxides.
\newblock {\em Reports on Progress in Physics}, 82(1):012501, 2018.

\bibitem{pai2018physics}
Yun-Yi Pai, Anthony Tylan-Tyler, Patrick Irvin, and Jeremy Levy.
\newblock {Physics of SrTiO$_3$-based heterostructures and nanostructures: a
  review}.
\newblock {\em Reports on Progress in Physics}, 81(3):036503, 2018.

\bibitem{vivek_normal}
M.~O.~Goerbig M.~Vivek and M.~Gabay.
\newblock {Topological states at the (001) surface of SrTiO$_3$}.
\newblock {\em Phys. Rev. B}, 95:165117, 2017.

\bibitem{scheurer2015topological}
Mathias~S Scheurer and J{\"o}rg Schmalian.
\newblock Topological superconductivity and unconventional pairing in oxide
  interfaces.
\newblock {\em Nature communications}, 6(1):1--10, 2015.

\bibitem{mohanta2014topological}
N~Mohanta and A~Taraphder.
\newblock {Topological superconductivity and Majorana bound states at the
  LaAlO$_3$/SrTiO$_3$ interface}.
\newblock {\em EPL (Europhysics Letters)}, 108(6):60001, 2014.

\bibitem{loder2015route}
Florian Loder, Arno~P Kampf, and Thilo Kopp.
\newblock Route to topological superconductivity via magnetic field rotation.
\newblock {\em Scientific reports}, 5:15302, 2015.

\bibitem{fukaya2018interorbital}
Yuri Fukaya, Shun Tamura, Keiji Yada, Yukio Tanaka, Paola Gentile, and Mario
  Cuoco.
\newblock Interorbital topological superconductivity in spin-orbit coupled
  superconductors with inversion symmetry breaking.
\newblock {\em Physical Review B}, 97(17):174522, 2018.

\bibitem{fidkowski2011majorana}
Lukasz Fidkowski, Roman~M Lutchyn, Chetan Nayak, and Matthew~PA Fisher.
\newblock Majorana zero modes in one-dimensional quantum wires without
  long-ranged superconducting order.
\newblock {\em Physical Review B}, 84(19):195436, 2011.

\bibitem{fidkowski2013magnetic}
Lukasz Fidkowski, Hong-Chen Jiang, Roman~M Lutchyn, and Chetan Nayak.
\newblock {Magnetic and superconducting ordering in one-dimensional
  nanostructures at the LaAlO$_3$/SrTiO$_3$ interface}.
\newblock {\em Physical Review B}, 87(1):014436, 2013.

\bibitem{mazziotti2018majorana}
Maria~Vittoria Mazziotti, Niccol{\`o} Scopigno, Marco Grilli, and Sergio
  Caprara.
\newblock {Majorana Fermions in One-Dimensional Structures at
  LaAlO$_3$/SrTiO$_3$ Oxide Interfaces}.
\newblock {\em Condensed Matter}, 3(4):37, 2018.

\bibitem{perroni2019evolution}
CA~Perroni, V~Cataudella, M~Salluzzo, M~Cuoco, and R~Citro.
\newblock Evolution of topological superconductivity by orbital-selective
  confinement in oxide nanowires.
\newblock {\em Physical Review B}, 100(9):094526, 2019.

\bibitem{perroni_ultimo}
J.~Settino, F.~Forte, C.~A. Perroni, V.~Cataudella, M.~Cuoco, and R.~Citro.
\newblock Spin-orbital polarization of majorana edge states in oxide nanowires.
\newblock {\em Phys. Rev. B}, 102:224508, 2020.

\bibitem{trier2019electric}
Felix Trier, Diogo~C Vaz, Pierre Bruneel, Paul No{\"e}l, Albert Fert, Laurent
  Vila, Jean-Philippe Attan{\'e}, Agn{\`e}s Barth{\'e}l{\'e}my, Marc Gabay,
  Henri Jaffr{\`e}s, et~al.
\newblock {Electric-field control of spin current generation and detection in
  ferromagnet-free SrTiO$_3$-based nanodevices}.
\newblock {\em Nano Letters}, 20(1):395--401, 2019.

\bibitem{Bibes_edelstein}
Annika Johansson, B\"orge G\"obel, J\"urgen Henk, Manuel Bibes, and Ingrid
  Mertig.
\newblock Spin and orbital edelstein effects in a two-dimensional electron gas:
  Theory and application to ${\mathrm{srtio}}_{3}$ interfaces.
\newblock {\em Phys. Rev. Research}, 3:013275, Mar 2021.

\bibitem{chirolli2022colossal}
Luca Chirolli, Maria~Teresa Mercaldo, Claudio Guarcello, Francesco Giazotto,
  and Mario Cuoco.
\newblock Colossal orbital edelstein effect in noncentrosymmetric
  superconductors.
\newblock {\em Phys. Rev. Lett.}, 128:217703, 2022.

\bibitem{levitov1985magnetoelectric}
LS~Levitov, Yu~V Nazarov, and GM~Eliashberg.
\newblock Magnetoelectric effects in conductors with mirror isomer symmetry.
\newblock {\em Soviet Journal of Experimental and Theoretical Physics},
  61(1):133, 1985.

\bibitem{revieworbitronic2021}
Dongwook Go, Daegeun Jo, Hyun-Woo Lee, Mathias Kl{\"a}ui, and Yuriy Mokrousov.
\newblock Orbitronics: Orbital currents in solids.
\newblock 135(3):37001, aug 2021.

\bibitem{chakhalian2020strongly}
Jak Chakhalian, Xiaoran Liu, and Gregory~A Fiete.
\newblock Strongly correlated and topological states in [111] grown transition
  metal oxide thin films and heterostructures.
\newblock {\em APL Materials}, 8(5):050904, 2020.

\bibitem{xiao2011interface}
Di~Xiao, Wenguang Zhu, Ying Ran, Naoto Nagaosa, and Satoshi Okamoto.
\newblock Interface engineering of quantum hall effects in digital transition
  metal oxide heterostructures.
\newblock {\em Nature Communications}, 2(1):1--8, 2011.

\bibitem{bruno2019band}
Flavio~Y Bruno, Siobhan McKeown~Walker, Sara Ricc{\`o}, Alberto De~La~Torre,
  Zhiming Wang, Anna Tamai, Timur~K Kim, Moritz Hoesch, Mohammad~S Bahramy, and
  Felix Baumberger.
\newblock Band structure and spin--orbital texture of the (111)-ktao3 2d
  electron gas.
\newblock {\em Advanced Electronic Materials}, 5(5):1800860, 2019.

\bibitem{boudjada2017magnetic}
Nazim Boudjada, Gideon Wachtel, and Arun Paramekanti.
\newblock Magnetic and nematic orders of the two-dimensional electron gas at
  oxide (111) surfaces and interfaces.
\newblock {\em Physical Review Letters}, 120(8):086802, 2018.

\bibitem{rout2017six}
PK~Rout, I~Agireen, E~Maniv, M~Goldstein, and Y~Dagan.
\newblock {Six-fold crystalline anisotropic magnetoresistance in the (111)
  LaAlO$_3$/SrTiO$_3$ oxide interface}.
\newblock {\em Physical Review B}, 95(24):241107, 2017.

\bibitem{monteiro2017two}
AMRVL Monteiro, DJ~Groenendijk, Inge Groen, Joeri de~Bruijckere, Rocco
  Gaudenzi, HSJ Van Der~Zant, and AD~Caviglia.
\newblock {Two-dimensional superconductivity at the (111) LaAlO$_3$/SrTiO$_3$
  interface}.
\newblock {\em Physical Review B}, 96(2):020504, 2017.

\bibitem{davis2017magnetoresistance}
S~Davis, Z~Huang, K~Han, T~Venkatesan, V~Chandrasekhar, et~al.
\newblock {Magnetoresistance in the superconducting state at the (111)
  LaAlO$_3$/SrTiO$_3$ interface}.
\newblock {\em Physical Review B}, 96(13):134502, 2017.

\bibitem{doennig2013massive}
David Doennig, Warren~E Pickett, and Rossitza Pentcheva.
\newblock {Massive Symmetry Breaking in LaAlO$_3$/SrTiO$_3$ (111) Quantum
  Wells: A Three-Orbital Strongly Correlated Generalization of Graphene}.
\newblock {\em Physical Review Letters}, 111(12):126804, 2013.

\bibitem{khanna2019symmetry}
Udit Khanna, Prasanna~K Rout, Michael Mograbi, Gal Tuvia, Inge Leermakers, Uli
  Zeitler, Yoram Dagan, and Moshe Goldstein.
\newblock {Symmetry and Correlation Effects on Band Structure Explain the
  Anomalous Transport Properties of (111) LaAlO$_3$/SrTiO$_3$}.
\newblock {\em Physical review letters}, 123(3):036805, 2019.

\bibitem{he2018observation}
Pan He, S~McKeown Walker, Steven S-L Zhang, Flavio~Yair Bruno, MS~Bahramy,
  Jong~Min Lee, Rajagopalan Ramaswamy, Kaiming Cai, Olle Heinonen, Giovanni
  Vignale, et~al.
\newblock {Observation of out-of-plane spin texture in a SrTiO$_3$ (111)
  two-dimensional electron gas}.
\newblock {\em Physical review letters}, 120(26):266802, 2018.

\bibitem{trama2021straininduced}
M.~Trama, V.~Cataudella, and C.~A. Perroni.
\newblock {Strain-induced topological phase transition at (111) SrTiO$_3$-based
  heterostructures}.
\newblock {\em Phys. Rev. Research}, 3:043038, Oct 2021.

\bibitem{trama2022gate}
Mattia Trama, Carmine~Antonio Perroni, Vittorio Cataudella, Francesco Romeo,
  and Roberta Citro.
\newblock {Gate tunable anomalous Hall effect at (111) LaAlO$_3 $/SrTiO$_3 $
  interface}.
\newblock {\em arXiv preprint arXiv:2202.04664}, 2022.

\bibitem{Keppler1998}
Hans Keppler.
\newblock {\em Crystal field theory}, pages 118--120.
\newblock Springer Netherlands, Dordrecht, 1998.

\bibitem{monteiro2019band}
AMRVL Monteiro, M~Vivek, DJ~Groenendijk, P~Bruneel, I~Leermakers, U~Zeitler,
  M~Gabay, and AD~Caviglia.
\newblock {Band inversion driven by electronic correlations at the (111)
  LaAlO$_3$/SrTiO$_3$ interface}.
\newblock {\em Physical Review B}, 99(20):201102, 2019.

\bibitem{de2018symmetry}
GM~De~Luca, R~Di~Capua, E~Di~Gennaro, A~Sambri, F~Miletto Granozio,
  G~Ghiringhelli, D~Betto, C~Piamonteze, NB~Brookes, and M~Salluzzo.
\newblock {Symmetry breaking at the (111) interfaces of SrTiO$_3$ hosting a
  two-dimensional electron system}.
\newblock {\em Physical Review B}, 98(11):115143, 2018.

\bibitem{notaklim}
{The comparison between the exact microscopic model~(\ref{hamiltonian_comp})
  and the effective Hamiltonian~(\ref{lsmodel}) breaks down for
  $|k|>0.5$~\cite{trama2022gate}.}

\bibitem{shen2014}
Ka~Shen, G.~Vignale, and R.~Raimondi.
\newblock Microscopic theory of the inverse edelstein effect.
\newblock {\em Phys. Rev. Lett.}, 112:096601, Mar 2014.

\bibitem{khan2017high}
Tahira Khan, Hui Zhang, Hongrui Zhang, Xi~Yan, Deshun Hong, Furong Han, Yuansha
  Chen, Baogen Shen, and Jirong Sun.
\newblock {High mobility 2-dimensional electron gas at LaAlO$_3$/SrTiO$_3$
  interface prepared by spin coating chemical methods}.
\newblock {\em Nanotechnology}, 28(43):435701, 2017.

\bibitem{bhowal2021orbital}
Sayantika Bhowal and Giovanni Vignale.
\newblock Orbital hall effect as an alternative to valley hall effect in gapped
  graphene.
\newblock {\em Physical Review B}, 103(19):195309, 2021.

\bibitem{bareille2014two}
C~Bareille, F~Fortuna, TC~R{\"o}del, F~Bertran, M~Gabay, O~Hijano Cubelos,
  A~Taleb-Ibrahimi, P~Le~Fevre, M~Bibes, A~Barth{\'e}l{\'e}my, et~al.
\newblock {Two-dimensional electron gas with six-fold symmetry at the (111)
  surface of KTaO$_3$}.
\newblock {\em Scientific reports}, 4(1):1--5, 2014.

\bibitem{vicente2021spin}
Luis~M Vicente-Arche, Julien Br{\'e}hin, Sara Varotto, Maxen Cosset-Cheneau,
  Srijani Mallik, Rapha{\"e}l Salazar, Paul No{\"e}l, Diogo~C Vaz, Felix Trier,
  Suvam Bhattacharya, et~al.
\newblock {Spin--charge interconversion in KTaO$_3$ 2D electron gases}.
\newblock {\em Advanced Materials}, 33(43):2102102, 2021.

\bibitem{khomskii2014transition}
Daniel Khomskii.
\newblock {\em Transition metal compounds}.
\newblock Cambridge University Press, 2014.

\bibitem{kalabukhov2007effect}
Alexey Kalabukhov, Robert Gunnarsson, Johan B{\"o}rjesson, Eva Olsson, Tord
  Claeson, and Dag Winkler.
\newblock {Effect of oxygen vacancies in the SrTiO$_3$ substrate on the
  electrical properties of the LaAlO$_3$/ SrTiO$_3$ interface}.
\newblock {\em Physical Review B}, 75(12):121404, 2007.

\bibitem{li2011formation}
Yun Li, Sutassana~Na Phattalung, Sukit Limpijumnong, Jiyeon Kim, and Jaejun Yu.
\newblock {Formation of oxygen vacancies and charge carriers induced in the
  n-type interface of a LaAlO$_3$ overlayer on SrTiO$_3$ (001)}.
\newblock {\em Physical Review B}, 84(24):245307, 2011.

\bibitem{walker2014control}
S~McKeown Walker, Alberto De~La~Torre, Flavio~Yair Bruno, Anna Tamai, TK~Kim,
  M~Hoesch, M~Shi, MS~Bahramy, PDC King, and F{\'e}lix Baumberger.
\newblock {Control of a two-dimensional electron gas on SrTiO$_3$ (111) by
  atomic oxygen}.
\newblock {\em Physical Review Letters}, 113(17):177601, 2014.

\end{thebibliography}

\end{document}